\documentclass[preprintnumbers,showkeys,superscriptaddress,showpacs,
nofootinbib,byrevtex,fleqn,prd,notitlepage]{revtex4-1}
\usepackage{amsmath,amsfonts,amssymb,amscd,amsxtra,amsthm}
\usepackage{graphicx}
\usepackage{bm,bbold}
\usepackage{epstopdf}
\usepackage{multirow}
\usepackage{hyperref}
\usepackage[fleqn]{nccmath}
\usepackage{empheq}
\usepackage{diagbox}
\input{epsf}
\newcommand{\be}{\begin{eqnarray}}
\newcommand{\ee}{\end{eqnarray}}
\newcommand{\ba}{\begin{array}}
\newcommand{\ea}{\end{array}}

\newcommand{\Slash}[1]{\ooalign{\hfil/\hfil\crcr$#1$}}
\usepackage[normalem]{ulem} 
\usepackage[dvipsnames]{xcolor} 
\renewcommand\sout{\bgroup \color{red} \ULdepth=-.5ex \ULset}

\begin{document}
\title{Nucleon quasi-Parton Distributions in the large N$_c$ limit}
%


\author{Hyeon-Dong Son}
\email[ E-mail: ]{Hyeon-Dong.Son@ruhr-uni-bochum.de}
\affiliation{Ruhr-Universit\"at Bochum, Fakult\"at f\"ur Physik und Astronomie,
Institut f\"ur Theoretische Physik II, D-44780 Bochum, Germany}
\author{Asli Tandogan}
\email[E-mail: ]{Asli.TandoganKunkel@ruhr-uni-bochum.de}
\affiliation{Ruhr-Universit\"at Bochum, Fakult\"at f\"ur Physik und Astronomie,
Institut f\"ur Theoretische Physik II, D-44780 Bochum, Germany}
\author{Maxim V. Polyakov}
\email[E-mail: ]{Maxim.Polyakov@tp2.ruhr-uni-bochum.de}
\affiliation{Petersburg Nuclear Physics Institute, Gatchina, 188300, St. Petersburg, Russia}
\affiliation{Ruhr-Universit\"at Bochum, Fakult\"at f\"ur Physik und Astronomie,
Institut f\"ur Theoretische Physik II, D-44780 Bochum, Germany}

\begin{abstract}
\noindent
In this letter, we investigate the nucleon quasi-parton distribution
functions in the chiral quark soliton model. We derive a set of
sum-rules depending on the velocity of the nucleon and
on the Dirac matrix defining the distribution functions.
We present numerical results for the isosinglet unpolarized
 distribution, in which we find that the anti-quark distribution breaks
 the positivity condition at nucleon velocities  of $v\approx 0.99\;(P_N\approx 7.0 M_N)$ and smaller.
 We found that, for the isosinglet unpolarized case, a large nucleon momentum
is required for the quasi-parton distribution to get close
enough to the usual parton distribution function.
\end{abstract}
\maketitle

\section{\normalsize \bf Introduction}

The parton distribution functions(PDFs) are
of great importance as they provide information
about the underlying structure of the hadrons.
In 2013 Xiangdong Ji \cite{Ji:2013dva} suggested that one can approach  the
PDFs asymptotically from the Euclidean region,
starting from nucleon matrix elements of {bilinear quark (gluon) operators with fields
separated by space-like distance (the quasi-parton distribution function (quasi-PDF))}, and then boosting the nucleon state to large momentum.
It has been followed by numerous researches to elaborate the
idea and  important lattice results have been obtained
\cite{Lin:2014zya,Alexandrou:2015rja,Chen:2016utp,Alexandrou:2016jqi,Constantinou:2017sej,Alexandrou:2017huk,Alexandrou:2018pbm,Alexandrou:2018eet,Lin:2018pvv,Alexandrou:2019lfo}.
 The detailed status on both the theory and the lattice simulations
 can be found in the recent review \cite{Cichy:2018mum} and in the community white papers, Refs. \cite{Lin:2017snn,Lin:2020rut}.

Switching our view point from the lattice,
one may utilise the chiral effective models
to provide the initial values of the QCD evolution for the
(quasi-)PDFs at a low renormalization point. For the model calculations
which were recently made in the context of the quasi-PDFs, we refer to
\cite{Gamberg:2014zwa,Broniowski:2017gfp,Bhattacharya:2019cme,Bhattacharya:2018zxi}.

A sound model calculation of the PDFs {has to} fulfil
 the key criteria such as the sum-rules and the positivity.
 Diakonov \textit{et al.} \cite{Diakonov:1996sr,Diakonov:1997vc}
achieved such satisfactory description of the nucleon PDFs
adopting the chiral quark soliton model \cite{Diakonov:1987ty}.
Here, we closely follow those works
where the authors already  in year 1997 introduced the object identical to
the quasi-distributions but concentrated on
the limiting case to study the usual nucleon PDFs \cite{Diakonov:1997vc}.
Here we simply compute them for the case where the nucleon
has finite momentum to investigate the properties of the quasi-PDFs
such as the sum-rules and the momentum evolutions.

In this letter,
we focus on the leading-twist quasi-PDFs, the isosinglet unpolarized
$u(x,v)+d(x,v)$ and the isovector polarized $\Delta u(x,v) - \Delta d(x,v)$ distributions
with the nucleon velocity $v$\footnote{ Instead of the nucleon momentum $P^z$, we take the nucleon
velocity $v$ as the parameter describing approach to the light-cone.
We find it much simpler and transparent to express our formulae and to discuss
the results. Conversion can be easily made with help of equation: $P^z= M_N v/\sqrt{1-v^2}$.}
 and the corresponding ones for the anti-quarks.
They are the {leading} components in the large $N_c$ approximation,  and
of particular interest as being related to the fundamental sum-rules.

Firstly, we dedicate a section to sketch the model. Next we
 derive the model expressions of the quasi-PDFs from the definitions of
the quark and anti-quark quasi-densities.
In the following section we calculate the first and next order Mellin moments and obtain the
 corresponding sum-rules.
After deriving the interpolation formula for the quasi-PDFs
 we present the numerical results on the isosinglet unpolarized
quark and anti-quark distributions and discuss their characteristics.
Finally, we come to the conclusion where we summarize the present work and
provide future perspectives.

\section{\normalsize \bf Nucleons as a soliton in the large $N_c$ limit}

We begin by describing the model framework briefly.
The detailed procedure and the formulae can be found in
the original paper Ref. \cite{Diakonov:1987ty} or in reviews Refs. \cite{Christov:1995vm,Diakonov:1997sj}.

Our starting point is the following effective action in the
large $N_c$,
\begin{align}\label{eq:eff_action_in_Nc}
\exp\left(i S_{\mathrm{eff}}[\pi(x)]  \right)
 = \int D\psi D\bar{\psi} \exp \left(
i \int d^4x \bar{\psi}(i \Slash{\partial} - M U^{\gamma_5}) \psi
 \right),\\
 U^{\gamma_5}(x) = \frac{1+\gamma_5}{2} U(x) + \frac{1-\gamma_5}{2} U(x)^\dagger,
\quad U(x) = \exp (i \pi^a(x) \tau^a ).
\end{align}
Here, $\psi$ is the quark field and $M$ is the dynamical quark mass, which is, in general,
momentum dependent.
Note that the above expression can be derived form the QCD at low renormalization
point where the vacuum is dominated by the instanton configurations \cite{Diakonov:1985eg,Diakonov:1983hh}.
In such a picture, the model scale is naturally given by the inverse of
the average instanton size $1/\bar{\rho} \sim 600 \;\mathrm{MeV}$ and
the dynamical quark mass is momentum dependent.
It is a reasonable approximation to switch off such momentum dependence
by assuming the low quark virtuality.
This makes the model calculations dramatically easier with still reasonable results
 in most cases, we simply need to introduce an artificial regularization scheme as a payoff.

We introduce the hedgehog ansatz for the pion field,
\begin{align}\label{eq:hedgehog}
 \pi^a(x) \tau^a = \frac{x^a \tau^a}{|x|} P(r).
\end{align}
$P(r)$ depends only on the distance $r=|\vec x|$
and is called the pion profile function.
Introducing the ansatz, we treat the quark-pion interaction
by a mean-field approach.
The Dirac equation can be solved to obtain the quark spectra
\begin{align}\label{eq:hamiltonian_equation}
H \Phi_n(\vec x, t) = E_n  \Phi_n(\vec x, t),
\end{align}
with the Dirac Hamiltonian,
\begin{align}\label{eq:dirac_hamiltonian}
H(U) = -i \gamma^0 \gamma^k \partial_k + M \gamma^0 U^{\gamma_5}.
\end{align}
Note that, the hedgehog ansatz \eqref{eq:hedgehog}
breaks the individual rotational symmetries
in the total angular momentum $J$ and the isospin $\tau$.
Instead we have the grandspin $K = J + \tau$ as good quantum number
as well as the parity $P=(-1)^{K,K+1}$, so called the hedgehog symmetry.
Among the Dirac spectra, there exists a distinct level with quantum number
$K^P = 0^+$ that emerges from the upper continuum as the pion mean-field
is turned on. We label this as the bound level.
The baryon number of the nucleon is given by the $N_c$ quarks
in the bound level. When the pion mean-field gets even stronger, eventually
the bound level falls into the negative continuum and the Skyrm picture of the
nucleon is applied.

By calculating the nucleon-nucleon correlation function and
passing it to the large Euclidean time, one can find the classical soliton energy,
\begin{align}\label{eq:classical_soliton_mass}
 M_{\mathrm{cl}} = N_c E_{\mathrm{level}} +N_c \sum_{occ} (E_n - E_n^0)^{reg}.
\end{align}
In the above {equation}, the first term corresponds to the contribution of the $N_c$ quarks in the bound level
 and the second term corresponds to that of the continuum part with
 the regularization ($reg$).
For the continuum part, the summation is over the negative (occupied) energy levels.
 One can write this in terms of
the summation over the upper continuum (non-occupied) levels
 by using the traceless nature of the Dirac Hamiltonian. We identify the classical
 soliton energy as the nucleon mass in this work, $M_N=M_{\mathrm{cl}}$.

\section{\normalsize \bf Nucleon quasi-parton distributions in the $\chi$QSM}

 Following Ref. \cite{Diakonov:1997vc}, we define the quark quasi-densities in the nucleon as\footnote{{The quasi-densities of Ref.~\cite{Diakonov:1997vc}
 coincide with quasi-distributions introduced afterwards in  Ref.~\cite{Ji:2013dva}}}
\begin{equation}
\label{eq:quasi-PDFs}
D_f(x,v)=\frac{1}{2 E_N}\int \frac{d^3 k}{(2\pi)^3}
\delta\left(x-\frac{k^3}{P_N}\right)\int d^3 x\ e^{-i {\bf k\cdot x}}
\langle N_v|  \bar \psi_f \left(-\frac{\bf{x}}{2},t\right) \Gamma
\psi_f\left(\frac{\bf{x}}{2},t\right)   |N_v\rangle,
\end{equation} and for the antiquarks

\begin{equation}
\label{eq:aquasi-PDFs}
\bar D_f(x,v)=\frac{1}{2 E_N}\int \frac{d^3 k}{(2\pi)^3}
 \delta\left(x-\frac{k^3}{P_N}\right)\int d^3 x
 \ e^{-i {\bf k\cdot x}} \langle N_v|  {\rm Tr}
 \left[
\Gamma \psi_f \left(-\frac{\bf{x}}{2},t\right)
\bar\psi_f\left(\frac{\bf{x}}{2},t\right) \right]  |N_v\rangle.
\end{equation}
In the above formulae the path ordered exponential is assumed {to guarantee the
gauge invariance of the nonlocal quark bilinear operator.}
 $E_N$ and $P_N$ are the energy and momentum of the nucleon moving with the velocity $v$
\begin{equation}
\nonumber
E_N=\frac{M_N}{\sqrt{1-v^2}}, \quad  P_N=\frac{M_N v}{\sqrt{1-v^2}},
\end{equation}
and $|N_v\rangle$ is the corresponding nucleon state.
{Here, the (anti-)quark momentum fraction $x$ extends from $0$ to $\infty$ and recover
the usual support from $0$ to $1$ in the PDF limit $v\to 1$.}
The spin matrix $\Gamma$ depends on the particular distribution one is interested in.
 For example, for the distribution of quarks polarized along or against
 the direction of the nucleon velocity one {can} use:
\begin{equation}
\nonumber
\Gamma=\gamma^0 \frac{1\pm \gamma_5}{2}\  {\rm or}\ \Gamma=\gamma^3 \frac{1\pm \gamma_5}{2},
\end{equation}
{or any of their linear combinations with proper normalization.}

%

{ In the large $N_c$ limit the nucleon matrix element entering Eqs.~(\ref{eq:quasi-PDFs},\ref{eq:aquasi-PDFs}) can be expressed in terms of
the
mean-field Green's function \cite{Diakonov:1997vc} which can be computed as:
\begin{align}\label{eq:greenftn}
\langle N_v | \mathrm{T} \left\{\psi(\vec{x}_1,t_1) \bar\psi(\vec{x}_2,t_2)
\right\}| N_v \rangle =& -S[\vec{v}] \left[\Theta(t_2 - t_1)
\sum_{occ} \Phi_n(\vec{x'}_1)\Phi_n^\dagger (\vec{x'}_2)
\gamma_0 \exp(-i E_n (t'_1 - t'_2)) \right. \cr
&\left.
-\Theta(t_1 - t_2) \sum_{nocc} \Phi_n(\vec{x'}_1)\Phi_n^\dagger (\vec{x'}_2)
\gamma_0 \exp(-i E_n (t'_1 - t'_2)) \right] S^{-1}[\vec{v}],\cr
\end{align}
where $\sum_{occ}$ and $\sum_{nocc}$ represent
the summation over the occupied and the non-occupied Dirac levels, respectively.
The $t'$ and $\vec{x'}$ are Lorentz transforms of the space-time coordinates:

\begin{equation}
\vec{x'}_{1,2}=\frac{\vec{x}_{1,2}-\vec{v} t_{1,2}}{\sqrt{1-v^2}}, \ \ t'_{1,2}=\frac{t_{1,2}+\vec{v}\cdot\vec{x}_{1,2}}{\sqrt{1-v^2}}.
\end{equation}
Eventually, $S[\vec{v}] $ is the Lorentz transformation matrix acting on the quark spinor indices:

\begin{equation}
S[\vec{v}] =\exp\left(\frac 14 [\gamma^0, \gamma^3] \omega\right), \ \ {\rm th}(\omega)=v.
\end{equation}
}
{Using Eq. \eqref{eq:greenftn} for Eqs.
 \eqref{eq:quasi-PDFs} and \eqref{eq:aquasi-PDFs} and representing the quark states
  in the momentum space,
 we obtain the expressions for the quasi-densities in the large $N_c$ limit:}

\begin{align}\label{eq:distrb_occ}
D_i(x,v) &= N_c M_N v \sum_{occ} \int \frac{d^3k}{(2\pi)^3} \delta(k^3 + v E_n - v M_N x)
\left[\Phi^\dagger_n(\vec{k}) (1 + v \gamma^0\gamma^3) \gamma_0 \Gamma_i \Phi_n (\vec{k})\right], \\
\label{eq:anti-distrb_occ}
\bar D_i(x,v) &= - N_c M_N v \sum_{occ} \int \frac{d^3k}{(2\pi)^3} \delta(k^3 + v E_n + v M_N x)
\left[\Phi^\dagger_n(\vec{k}) (1 + v \gamma^0\gamma^3) \gamma_0 \Gamma_i \Phi_n (\vec{k})\right],
\end{align}
{in terms of quark orbitals (\ref{eq:hamiltonian_equation}) in the meson mean-field. The index $i$ above combines the flavour and helicity indices of quarks. Using} the anti-commutativity of the quark fields {at equal time},
it is possible to represent the densities equivalently by the summation over the non-occupied Dirac levels,
\begin{align}\label{eq:distrb_nocc}
D_i(x,v) &= -N_c M_N v \sum_{nocc} \int \frac{d^3k}{(2\pi)^3} \delta(k^3 + v E_n - v M_N x)
\left[\Phi^\dagger_n(\vec{k}) (1 + v \gamma^0\gamma^3) \gamma_0 \Gamma_i \Phi_n (\vec{k})\right], \\
\bar D_i(x,v) &=  N_c M_N v \sum_{nocc} \int \frac{d^3k}{(2\pi)^3} \delta(k^3 + v E_n + v M_N x)
\left[\Phi^\dagger_n(\vec{k}) (1 + v \gamma^0\gamma^3) \gamma_0 \Gamma_i \Phi_n (\vec{k})\right].
\end{align}
In the above expressions, we traced out the colour space and obtained the overall factor of $N_c$.
Notice as well that the factor of nucleon mass $M_N$ is also order of $N_c$.
One has to keep in mind that it is necessary to take the vacuum subtraction
$-(H \to H_0)$ where $H_0$ is the free Dirac Hamiltonian. This will be consistently
omitted in the following formulae for brevity and will be mentioned if required.

Now one has to take certain combinations of Eqs. \eqref{eq:distrb_occ} and
\eqref{eq:anti-distrb_occ} to get the desired quasi-parton distribution functions.
{The leading distributions at large $N_c$ are isoscalar  unpolarized ($u+d$) and isovector polarized ($\Delta u-\Delta d$) ones.}
For the isosinglet unpolarized quasi-distribution
one has to take average over the flavour space and sum up the polarizations:

\begin{align}\label{eq:isup_distribution_occ}
\sum_f q_f(x,v) & = N_c M_N v \sum_{occ} \int \frac{d^3k}{(2\pi)^3}
\delta(k^3+ v E_n - v M_N x)
\Phi_n^\dagger(\vec k) (1 + v \gamma^0\gamma^3) \gamma^0 \Gamma \Phi_n(\vec k).
\end{align}
In this expression, we absorbed the anti-quark distribution using the identity
$\bar{q}(x,v) = - q(-x,v)$. Thus the variable $x$ ranges from negative to positive infinity.

For the isovector polarized quark and anti-quark quasi-distribution,
 we obtain the following with the identity $\Delta \bar{q}(x,v) = \Delta q(-x,v)$.
\begin{align}\label{eq:ivp_distribution_occ}
\Delta q_f (x,v) & = - \frac{2}{3}(T_3)_{ff} N_c M_N v \sum_{occ} \int \frac{d^3k}{(2\pi)^3}
\tau^3 \delta(k^3+ v E_n - v M_N x)
\Phi_n^\dagger(\vec k) (1 + v \gamma^0\gamma^3) \gamma^0 \Gamma \gamma_5 \Phi_n(\vec k).
\end{align}
Here, $T_3={\rm diag}(1/2, -1/2)$ is the proton isospin matrix.
Again, the above expressions can be equally represented as a summation over
 the non-occupied states with opposite sign.

The Dirac matrix $\Gamma$ in Eqs. \eqref{eq:isup_distribution_occ} and \eqref{eq:ivp_distribution_occ}
{is any linear combination of  $\gamma^0$ and $\gamma^3$}.
Note that there is no unique definition of the quasi-PDFs, {one can use any Dirac structures} as far as they
{provide} the correct limit to the usual PDFs. In the next section, we work with
the both $\gamma^0$ and $\gamma^3$ to examine the sum-rules.

\section{\normalsize \bf Sum-rules}
Discussing the sum-rules, it is more convenient to use the
frequency representation instead of the discrete summation using
the following identity
\begin{align}\label{eq:frequency}
\sum_{{occ}} |n \rangle \langle n | \exp (-i z^0 E_n) =\int^{E_{\mathrm{level}}+\epsilon}_{-\infty} dw \delta(w - H)
\exp (-i z^0 w).
\end{align}

The leading Mellin moment of the distribution \eqref{eq:distrb_occ} can be written as
\begin{align}\label{number_sumrule_proof1}
\int^\infty_{-\infty} dx\; \sum_{f} q(x,v)
&= N_c \int^{E_{\mathrm{level}}+\epsilon}_{-\infty} dw\;
\mathrm{Sp} [\delta(w-H) (1+v\gamma^0\gamma^3)\gamma^0\Gamma] - (H \to H^0 )\cr
&= N_c \; \mathrm{Sp} [\Theta(-H+E_{\mathrm{level}}+\epsilon) (1+v\gamma^0\gamma^3)\gamma^0\Gamma ]
- (H \to H^0 ).
\end{align}
One can show that the term proportional to $\gamma^0\gamma^3$ inside the trace
vanishes with the hedgehog symmetry. Then we are left with the following trace
\begin{align}\label{number_sumrule_proof2}
 \mathrm{Sp} [\Theta(-H+E_{\mathrm{level}}+\epsilon)-\Theta(-H_0)],
\end{align}
which is nothing but the baryon number $B$ as it counts the number of the quarks
occupying the negative energy levels and the bound level subtracted by that
of the negative levels with the vacuum Hamiltonian.
Thus we obtain
\begin{align}\label{eq:sumrule_B}
\int^\infty_0 dx \; ( q(x,v) - \bar{q}(x,v) ) &=
\begin{cases}
\;\; N_c B \quad  &\Gamma = \gamma^0\\
\;\;  v N_c B \quad  &\Gamma = \gamma^3 \; .
\end{cases}
\end{align}

Similarly, from the next-order Mellin moment, we obtain the momentum sum-rule:
\begin{align}\label{eq:sumrule_M2}
\int^\infty_0 dx \;x ( q(x,v) + \bar{q}(x,v) ) &=
\begin{cases}
\;\;1 \quad  &\Gamma = \gamma^0\\
\;\;v  \quad  &\Gamma = \gamma^3 \; .
\end{cases}
\end{align}
Deriving the momentum sum-rule we use the identity
$\mathrm{Sp}[\Theta(-H+E_{\mathrm{level}}+\epsilon)\gamma^0\gamma^3P^3]=0$,
which can be proven using the saddle point equation of the effective action \cite{Diakonov:1996sr}.
In that case, the chiral field should minimize the energy functional
and the sum-rule is only satisfied strictly with such solution.
Note the nucleon momentum is soley carried by the quarks
 being the only effective degrees of freedom in the model picture.

Finally, for the isovector polarized distribution, we obtain the
Bjorken sum-rule,
\begin{align}\label{eq:sumrule_ga}
\int^\infty_0 dx \; ( \Delta u(x,v)- \Delta d(x,v)
 + \Delta \bar{u}(x,v)- \Delta \bar{d}(x,v) ) &=
\begin{cases}
\;\; v g_A \quad  &\Gamma = \gamma^0 \\
\;\;\;\; g_A \quad  &\Gamma = \gamma^3 \; ,
\end{cases}
\end{align}
where the nucleon axial charge $g_A$ has the following expression
\begin{align}\label{eq:nucleon_axial_charge}
g_A = - \frac{N_c}{3} \int^{E_{\mathrm{level}}+\epsilon}_{-\infty} dw\; \mathrm{Sp}
[\delta(w-H) \tau^3 \gamma^0 \gamma^3 \gamma_5] - (H \to H_0).
\end{align}
Eqs. \eqref{eq:sumrule_B},\eqref{eq:sumrule_M2} and \eqref{eq:sumrule_ga} are
the generalized version of the usual baryon number-, momentum-, and
Bjorken sum-rules of the nucleon. Note that two choices of the
Dirac matrix $\Gamma= \gamma^0$ and $\gamma^3$ results in different
velocity dependence for the sum-rules. Clearly, in the limit $v\to1$
they become the usual sum-rules.

Although the proofs here are given within the model framework, the sum-rules
 can be understood from more general point of view.
 Taking the Mellin moments, the non-locality of the quark bilinear in the matrix elements is
lifted and in general we are left with the charges of the symmetry currents.
For example, the zeroth component of the vector current is just the number density.
With the conservation of the vector current, the spatial component($\gamma^3$) should
be proportional to the `velocity' by virtue of the continuity.
For the momentum sum-rule, different choices of the operator $\Gamma = \gamma^0 $ and $ \gamma^3$
 correspond to taking different components of the energy-momentum tensor
  matrix element, $\sim T^{30}$ and $T^{33}$, respectively.
 {In general case the momentum sum-rule (\ref{eq:sumrule_M2}) has the following form (see, e.g. \cite{Bhattacharya:2019cme}):

  \begin{align}\label{eq:sumrule_M2_general}
\int^\infty_0 dx \;x ( q(x,v) + \bar{q}(x,v) ) &=
\begin{cases}
\;\; M_2^q \quad  &\Gamma = \gamma^0\\
\;\;v M_2^q -\bar c^q(0) \frac{1-v^2}{v} \quad  &\Gamma = \gamma^3 \; .
\end{cases}
\end{align}
 where $M_2^q$ is the momentum fraction carried by the quarks in the nucleon,
$\bar c^q(t)$ is the form factor of the energy-momentum tensor.\footnote{For its definition and discussion of its properties see, e.g. Ref.~\cite{Polyakov:2018exb}}
The form factor $\bar c^q(t)$ is zero
 in the chiral quark-soliton model (as it appears in the next to leading order in the instanton density, see \cite{Polyakov:2018exb}), also $M_2^q=1$ in the model.
 Taking this into account, one can reproduce Eq.~(\ref{eq:sumrule_M2}) in the chiral quark-soliton model.
  }

  In the case of the Bjorken sum-rule,
  $\Gamma=\gamma^3$ corresponds to the third component of the nucleon spin $S^3$ times
  the axial charge $g_A$ whereas the $\Gamma=\gamma^0$ case can be
   related to $\vec{S} \cdot \vec{v}g_A$.

\section{\normalsize \bf Quasi-parton distributions in terms of the pion mean-field}
While the quasi-PDFs \eqref{eq:isup_distribution_occ} and \eqref{eq:ivp_distribution_occ}
can be computed directly by calculating the Dirac spectra, it often
takes significant computational time and techniques.
Here we suggest taking further approximation which makes
the numerical computation pretty simple and
is feasible to inspect the structure of the divergences.

One can expand the real part of the effective action \eqref{eq:eff_action_in_Nc},
 taking $pM(U-1)/(p^2+M^2)$ as the small expansion parameter where $p$ is the
 characteristic pion momentum.
Such expansion is valid for three limiting cases: when the pion field is small,
 when the pion momentum is small $p\ll M$ , and when the pion momentum is large $p\gg M$.
 This is called the interpolation formula \cite{Diakonov:1987ty}
and provides a reasonable result
 compared to the full calculation even only
 when the lowest order is considered, for example look Ref. \cite{Diakonov:1996sr}.

Accordingly, the isosinglet unpolarized distribution can be expanded as below
\begin{align}\label{eq:expansion}
\sum_f q_f(x,v) = \sum^\infty_{m=0} \left[\sum_f q_f(x,v) \right]^{(m)},
\end{align}
with
\begin{align}\label{eq:expansion-n}
\left[\sum_f q_f(x,v) \right]^{(m)}=
\frac{N_c M_N}{\mathcal{T}} \mathrm{Im} \; \mathrm{Tr}
\bigg[  (i \Slash{\partial} + M U^{-\gamma_5})
(-1)^m \left[(-\partial^2 -M^2 +i\epsilon)^{-1}
iM(\Slash{\partial}U^{-\gamma_5} )\right]^m\cr
 (-\partial^2 -M^2 +i\epsilon)^{-1}
\delta(i \Slash{n} - x M_N) \Slash{\bar n} \bigg] - (U \to 1),
\end{align}
where $\mathrm{Tr}$ is the functional trace and $\mathcal{T}$ denotes the Euclidean time {which drops in the final result}.
Here we defined the quasi-light vectors for brevity
\begin{align}\label{eq:nbar_n}
\bar n &=(1,0,0, -v), \\
 n &=(1,0,0,-1/v).
\end{align}

The $m=0$ term in Eq. \eqref{eq:expansion-n} is trivially zero.
The leading order $m=1$ can be written as follows in the momentum representation
\begin{align}\label{eq:expansion-n=1_mom}
\left[\sum_f q_f(x,v) \right]^{(1)}
= - N_c M_N M^2 \;\mathrm{Im} \int \frac{d^3 k}{(2\pi)^3} \frac{d^4 p}{(2\pi)^4}
((p+k)^2-M^2 + i\epsilon )^{-1} (p^2-M^2+i\epsilon)^{-1} \cr
\delta ( n \cdot p - x M_N) ( \bar n \cdot k)
\; \mathrm{Sp} \big[\tilde U(\vec k)^\dagger\tilde U(\vec k) \big],
\end{align}
where $k^\mu=(0, \vec k)$ and the Fourier transform
of the pion mean-field is defined as follows
\begin{equation}\label{eq:utilde}
\tilde U(\vec k) = \int d^3x e^{-i \vec k \cdot \vec x} \big[U(\vec x) -1\big].
\end{equation}
To calculate the momentum $p$-integral, we {use the Sudakov} decomposition,
\begin{align}\label{eq:integral_light}
\int \frac{d^4p}{(2\pi)^4}
= \frac{1}{2} \int \frac{dp^+}{2\pi}\frac{dp^-}{2\pi}\frac{d^2p_\perp}{(2\pi)^2}.
\end{align}
We first integrate over $p^+$ using the $\delta$-function
and then perform the $p^-$ integral. We present a detailed analysis on the poles in $p^- $in Appendix
 \ref{appendix1}.
We check that the quark-loop momentum integral and the limit $v\to 1$
 are commutable, i.e. correct limit of the quasi-PDFs to PDFs is achieved.
 We also observe that the remaining integration over $\vec p_\perp$ is logarithmically
divergent. In Refs. \cite{Diakonov:1996sr,Diakonov:1997vc} the authors introduced the Pauli-Villars
as one of the `good' regularization method preserving the required properties of the PDFs,
such as the positivity and sum-rules.
For the numerical calculation of the quasi-PDFs, we follow the same strategy:
 the Pauli-Villars regularization with single subtraction.

\section{\normalsize \bf Numerical results and discussions}
For the numerical results,
 we focus only on the isosinglet unpolarized quasi-distribution
$u(x,v)+d(x,v)$ with $\Gamma=\gamma^0$.
While the bound level contribution
is calculated exactly by solving the Dirac Hamiltonian,
we use the interpolation formula \eqref{eq:expansion} and \eqref{eq:expansion-n}
up to $m=2$ for the continuum part. The full calculation including the isovector polarized with
the both Dirac matrices $\Gamma=\gamma^0$ and $\Gamma=\gamma^3$ will be
covered in another publication.

For simplicity in the computation procedure, we use the following ansatz for
the pion mean-field
\begin{align}\label{eq:pion_arctan}
P(r) = - 2\; {\rm arctg} \left(\frac{r_0^2}{r^2} \right),
\end{align}
with $r_0 \approx 1.0/M$. This choice of the pion mean-field
has been used to successfully describe various nucleon observables and
typical deviation on the results between using the ansatz (\ref{eq:pion_arctan})
and the self-consistent solution which minimizes the energy functional
is known to be up to around $10\%$
\cite{Diakonov:1987ty,Diakonov:1996sr,Diakonov:1997vc,Diakonov:1997sj}.

As discussed in the previous section,
we adopt the Pauli-Villars scheme with
single subtraction to tame the logarithmic divergences in
the quark-loop momentum integral.
The Pauli-Villars mass $M_{PV}$ is determined by using the pion decay constant,
\begin{align}\label{eq:pv_mass}
F_\pi^2 = \frac{N_c M^2}{4\pi^2}\ln \frac{M_{PV}^2}{M^2}.
\end{align}
Using $F_\pi = 93 ~\mathrm{MeV}$ and $M=350 ~\mathrm{MeV}$,
 we obtain $M_{PV}= 560 ~\mathrm{MeV}$.
At the same time, we obtain the nucleon mass $M_N \approx 1.15 ~\mathrm{GeV}$ which
will be used in the numerical analysis.

{To discuss the velocity (momentum) evolution of the qPDFs, we take rather wide range of the velocity values.
In Table \ref{table:v-P}, the conversion between the velocity $v$ and the factor $P_N/M_N$ is given.
 The nucleon velocities $v=0.7$ and $v=0.9$ corresponds to $P_N =M_N$
and $P_N=2 M_N$, respectively, in the typical momentum range used in the current lattice simulations.}

\begin{table}[ht]
\begin{tabular}{ c c c c c}
\hline\hline
$v$ 	  & $0.999$ & $0.99$ & $0.9$ & $0.7$ \tabularnewline
$P_N/M_N$ & $22$    & $7$    & $2$   & $1$ \tabularnewline
\hline\hline
\end{tabular}
\caption{Conversion between the nucleon velocity $v$ and the factor $P_N/M_N$. }
\label{table:v-P}
\end{table}

\begin{figure}[ht!]
\begin{center}
\includegraphics[width=7.5cm]{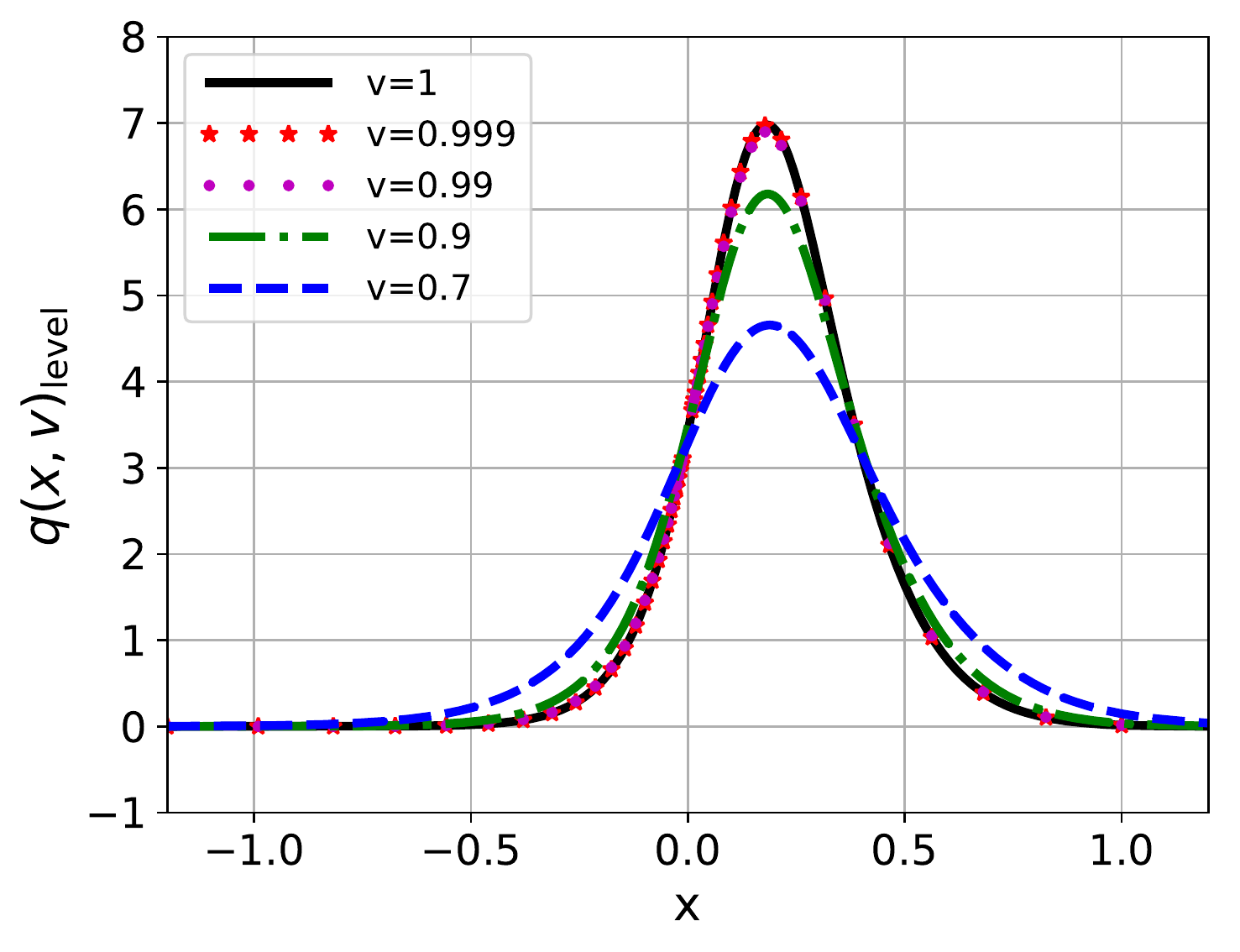}
\includegraphics[width=7.5cm]{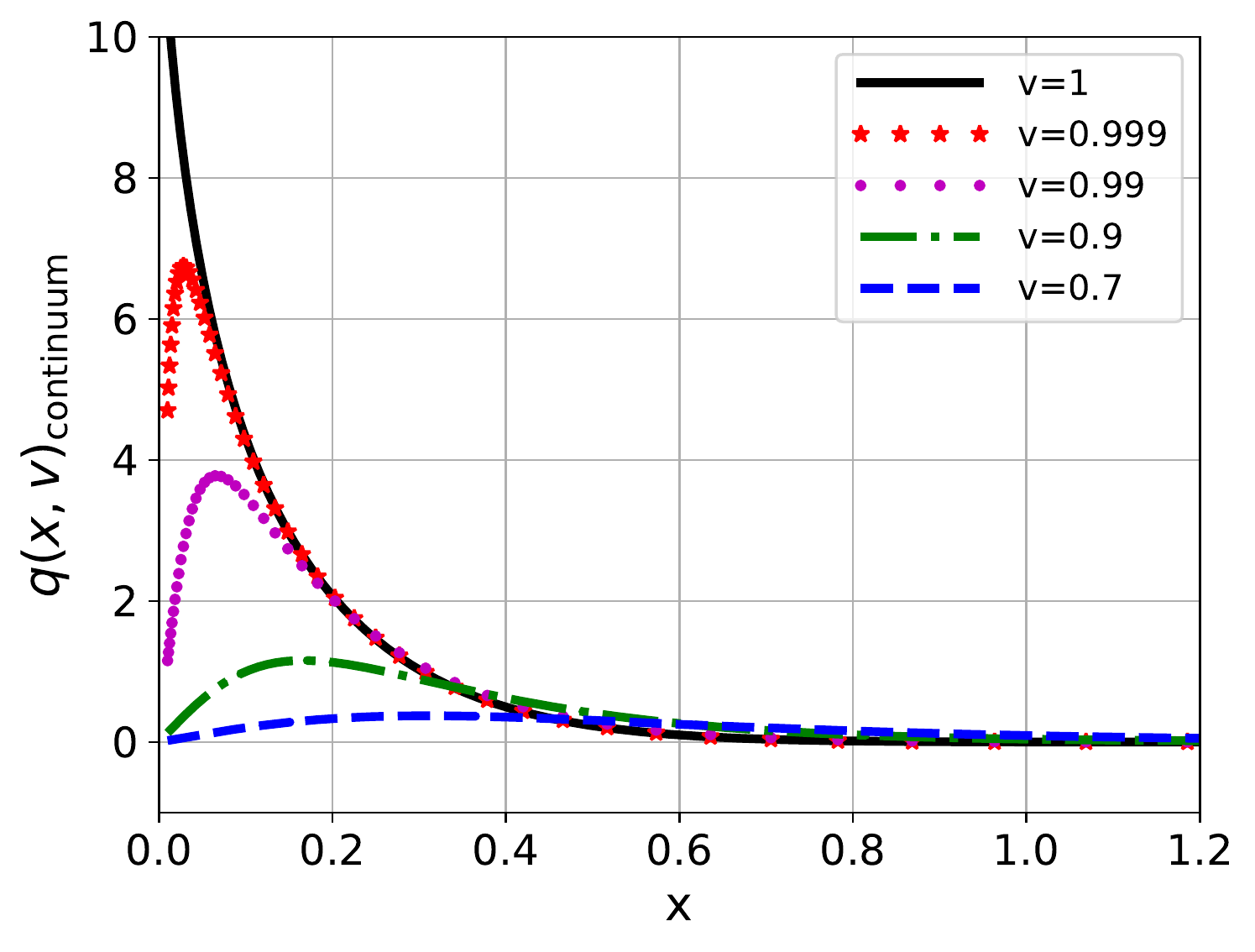}
\caption{The bound level (left) and continuum (right) parts of the isosinglet unpolarized distribution.
 For the continuum part, $\bar{q}(x,v)=q(x,v) $.}
\label{fig1}
\end{center}
\end{figure}

\begin{figure}[ht!]
\begin{center}
\includegraphics[width=7.5cm]{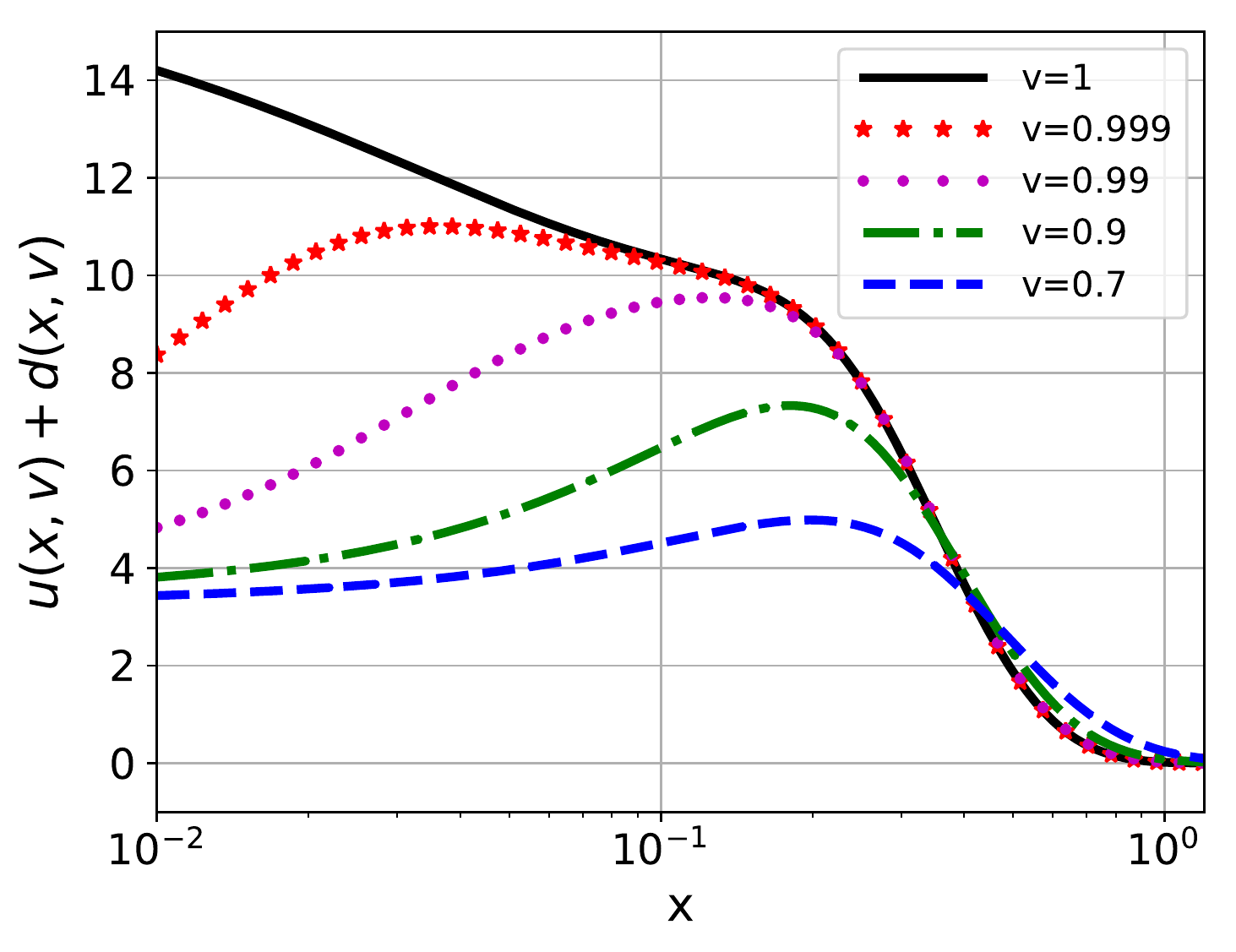}
\includegraphics[width=7.5cm]{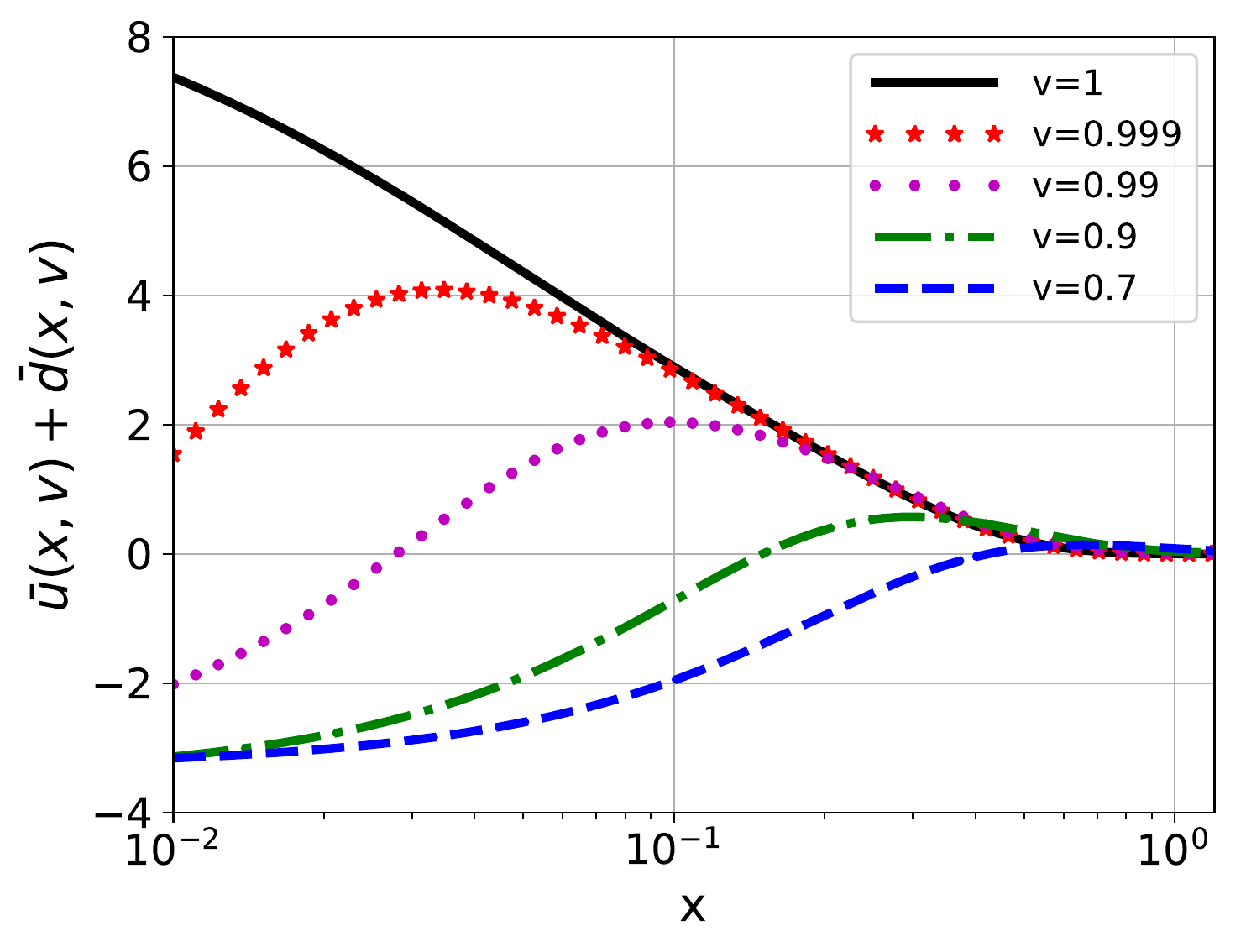}
\caption{The quark (left) and anti-quark (right) isosinglet unpolarized distributions}
\label{fig2}
\end{center}
\end{figure}

In Fig. \ref{fig1}, the bound level (left) and continuum (right) contributions to the isosinglet unpolarized
distribution are displayed for different $v$ values.
While the bound level part shows mild dependence on the nucleon velocity,
the continuum part changes significantly rapidly at small $x$ region.
Already for $v=0.999$, which corresponds to $P_N \approx 22 M_N$ in the nucleon momentum,
the distribution is a half of the $v=1$ distribution at $x=0$.
Such distinct behaviour of the bound and continuum contribution
leads to an interesting consequence. See Fig. \ref{fig2},
where the total quark (left) and anti-quark (right) distributions are shown.
Note that the anti-quark distributions become negative at the nucleon velocity
$v=0.99$ ($P_N \approx 7.0 M_N$) and smaller.
As the bound level negatively contributes to the anti-quark
distribution, the role of the positive
and sizable contribution from the continuum part is essential to
guarantee the positiveness for the usual PDFs \cite{Diakonov:1996sr,Diakonov:1997vc}.
In our case, the positiveness is not required
for the quasi-PDFs as the nucleon is off the light-cone where the
probability interpretation is not valid. Indeed we observe that the
positivity condition is not satisfied.
In Ref. \cite{Gamberg:2014zwa}, using the diquark spectator model,
 the authors also found a breakdown of the Soffer positivity condition
 for the quark quasi-PDFs.

We check numerically that the baryon number sum-rule is well satisfied.
On the other hand for the momentum sum-rule, we obtain $M_2 \approx 0.95$,
which is $5\%$ deviated from the correct value. The reason is simple:
we use the ansatz \eqref{eq:pion_arctan} which is not the
true solution which minimizes the action. Note that we derived the
momentum sum-rule using the equation of motion for the chiral field
$\delta_U S_{\rm eff} = 0$. Thus we expect that the sum-rule will be fulfilled
when we use the self-consistent pion mean-field \cite{Weiss:1997rt}.
 Nevertheless we check that the $v$-dependence of the both
 baryon number and momentum sum-rules \eqref{eq:sumrule_B} and \eqref{eq:sumrule_M2}
  are satisfied: they are independent on $v$ for $\Gamma = \gamma^0$.

\begin{figure}[ht!]
\begin{center}
\includegraphics[width=7.5cm]{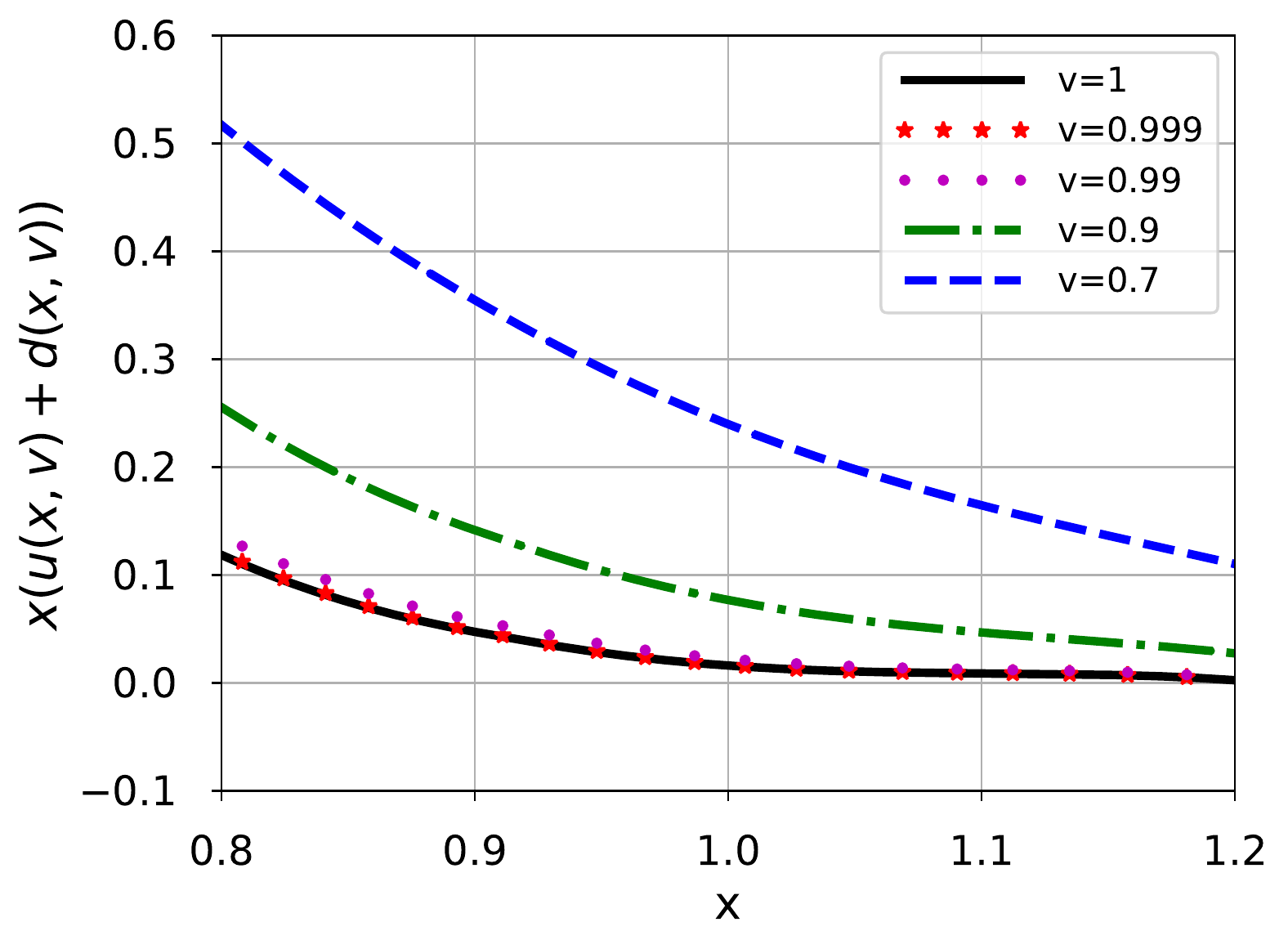}
\includegraphics[width=7.5cm]{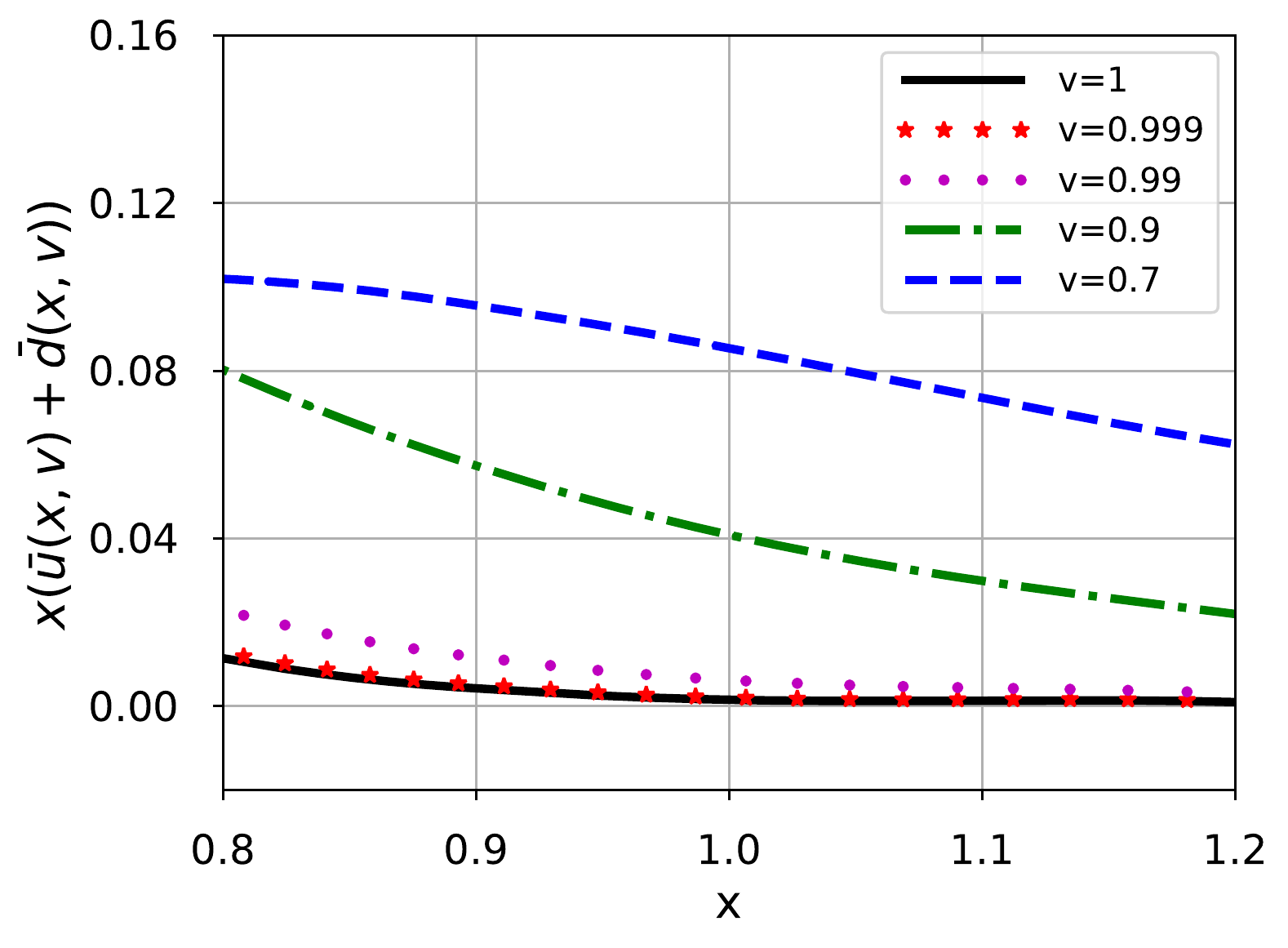}
\caption{ A magnified view of $x (u(x,v)+d(x,v) )$ (left) and $x (\bar{u}(x,v)+\bar{d}(x,v))$ (right), near $x=1$.
 }
\label{fig3}
\end{center}
\end{figure}

Several studies in framework of the diquark model \cite{Bhattacharya:2018zxi,Bhattacharya:2019cme} and studies of renormalon contribution
to quasi-PDFs \cite{Braun:2018brg} show that the quasi-PDFs differ strongly from PDFs in the vicinity of $x=1$.
{In Fig. \ref{fig3}, the quasi-PDFs $x q(x,v) $ and $x \bar{q}(x,v)$ are plotted from $x=0.8$ to $x=1.2$ to
provide a clearer view at large $x$. A significant discrepancy between the PDF and the quasi-PDFs at $v=0.9$ and $v=0.7$ is observed.
 The corresponding ratio quasi-PDF$(x)/{\rm PDF}(x)\sim \exp(const (1-v) N_c)$ is exponentially
large for $x\sim 1$ in the large $N_c$ limit
\footnote{In the framework used, the usual PDF ($v=1$) develops an exponential tail beyond the support bound $x=1$. This is due to the
fact that the nucleon is a heavy object $M_N = \mathcal{O}(N_c)$ in the large $N_c$ limit.}.

Let us discuss briefly the shape of the quasi-PDF in the context of the sum-rules.
In general, the quasi-PDFs become softer in smaller nucleon velocity, while satisfying the sum-rules (\ref{eq:sumrule_B}-\ref{eq:sumrule_ga}).
Hence the missing portion of the sum-rule within the usual interval $[0,1]$ of $x$ should be complemented by the extended tail ($x> 1$).
To measure this, one may define the following quantity
\begin{align}\label{eq:R_x0}
R^N(x_0,v)  = \frac{\int^{x_0}_0 dx\;x^{N-1} ( q(x,v) - (-1)^{N-1}\bar{q}(x,v) )}{ \int^{x_0}_0 dx\;x^{N-1} ( q(x,v=1) - (-1)^{N-1}\bar{q}(x,v=1))},
\end{align}
which is the ratio of the moments of the isoscalar unpolarized distributions but integrated over $x$ only up to a finite point $x=x_0$.
In the case of $N=1,2$, they are related to the ratio of the baryon number and the momentum sum-rule, respectively.
From Eqs. (\ref{eq:sumrule_B}) and (\ref{eq:sumrule_M2}), we expect the following limit
\begin{align}\label{eq:R_limit}
\lim_{x_0 \to \infty } R^{N=1,2}(x_0,v) =
\begin{cases}
\;\; 1 \quad  &\Gamma = \gamma^0 ,\\
\;\;  v \quad  &\Gamma = \gamma^3. \;
\end{cases}
\end{align}

In Table \ref{table:R_x0}, using our numerical results, we evaluate the ratio \eqref{eq:R_x0} with $N=1,2$ varying $x_0$ from
$x_0=0.8$ to $x_0=1.2$.
\begin{table}[ht]
 \begin{tabular}{ c c c c c c || c c c c c c  }
\hline\hline
$R^{N=1}(x_0,v)$  & & & & & & $R^{N=2}(x_0,v)$ & & & & & \tabularnewline
\hline\hline
\diagbox[innerwidth=1.5cm,height=1.2\line]{$v$}{$x_0$}   & $0.8$     & $0.9$     & $1.0$    & $1.1$  & $1.2$ & \diagbox[innerwidth=1.5cm,height=1.2\line]{$v$}{$x_0$}   & $0.8$     & $0.9$     & $1.0$    & $1.1$  & $1.2$ \tabularnewline
\hline
$0.9$ & $0.99$    & $1.00$    & $1.00$   & $1.00$ & $1.00$ & $0.9$ &$0.92$    & $0.94$    & $0.96$   & $0.97$ & $0.98$ \tabularnewline
$0.7$ & $0.97$    & $0.98$    & $0.99$   & $1.00$ & $1.00$ & $0.7$ &$0.76$    & $0.81$    & $0.85$   & $0.88$ & $0.90$\tabularnewline
\hline\hline
\end{tabular}
\caption{Numerical values of the ratio $R^N(x_0,v)$ given in Eq. \eqref{eq:R_x0}, for $N=1$ (left) and for $N=2$ (right). For $N=2$
and $v=0.7$, the ratio turns out to be only $0.9$ at $x_0=1.2$, compared to its limiting value $R^{N=2}(x_0=\infty,v)=1$.}
\label{table:R_x0}
\end{table}
We find that the ratio $R^{N=2}(x_0,v=0.7)$ is still underestimated at $x_0=1.2$
by $10\%$ when compared to its limiting value $R^{N=2}(x_0=\infty,v)=1$.
}

\section{\normalsize \bf Summary and outlook}
We demonstrated how to calculate the nucleon quasi-parton distribution
functions within the framework of the chiral quark soliton model.
We found the sum-rules for the isosinglet unpolarized and isovector polarized
distributions which generalize the usual ones for the nucleon PDFs.
 It is intriguing that the operators defining the matrix elements
$\Gamma=\gamma^0, \gamma^3$ exhibit different nucleon velocity
 dependencies of the sum-rules. Although they are derived in a model approach,
 we argued that they can be understood in the general context, taking the
 notion of the local QCD operators and their components.
Finally we discussed the numerical results on the quark and anti-quark
 isosinglet unpolarized distributions.
We observed that in particular, the continuum contribution at small $x$ has sharp dependence on $v$
which consquently leads the  anti-quark distribution to be \textit{negative} at $v \sim 0.99$
($P_N \approx 7.0 M_N$) and smaller.
We would like to spotlight the observation that the nucleon is required to have quite large momentum
so that its isosinglet unpolarized quasi-PDFs to be close enough to the usual PDFs,
due to the rapidly varying continuum contribution with respect to
the nucleon velocity.

We stress that the numerical test of the momentum sum-rule is around $5\%$ underestimated
 when using the  ansatz (\ref{eq:pion_arctan}).
  The sum-rule is satisfied only when the self-consistent mean-field is used.
To numerically test the momentum sum-rule with correct velocity dependencies
 for the both cases $\Gamma= \gamma^0$ and $\gamma^3$, it is indeed required to
 perform the computation using the self-consistent profile
  and the full calculation instead of using the interpolation formula.
Hence we plan to provide the improved results for both the polarized and unpolarized quasi-PDFs in a forthcoming work,
 with elaborated discussions on the sum-rules Eqs. (23-25) more in detail.
  In particular, the polarized distribution $\Delta u - \Delta d$ will be of interest as the corresponding
   lattice results can be found, for instance, in Refs.
   \cite{Alexandrou:2015rja,Chen:2016utp,Alexandrou:2016jqi,Alexandrou:2017huk,Alexandrou:2018pbm,Alexandrou:2018eet,Lin:2018pvv,Alexandrou:2019lfo}.

Apart from the practical usage of the quasi-PDFs on the lattice,
Radyushkin suggested that they have deeper theoretical ground
\cite{Radyushkin:2016hsy,Radyushkin:2017cyf}.
For instance, the quasi-PDF is
 related to the transverse momentum distributions(TMDs) {and Ioffe time pseudo-parton distribution functions
 \cite{Radyushkin:2017cyf,Broniowski:2017gfp}. These new ideas were already tested in the lattice simulations \cite{Orginos:2017kos,Joo:2019bzr,Joo:2019jct,Karpie:2018zaz}}
This will be an interesting future subject to look into carefully how the transformations
between various distributions are realised in the soliton picture of baryons.

\section*{\normalsize \bf Acknowledgements}
{We are grateful to Krzysztof Cichy, Martha Constantinou and Andreas Metz for correspondence.}
This work is supported by the DFG through the Sino-German CRC 110
 ``Symmetries and the Emergence of Structure in QCD" (Grant No. TRR110).

\appendix

\section{\normalsize \bf Determination of the poles}
\label{appendix1}
Let us write the $p$ integral in \eqref{eq:expansion-n=1_mom} as
\begin{align}
 I &\equiv \int \frac{d^4 p}{(2\pi)^4} \frac{1}{(p+k)^2 - M^2 +i \epsilon}
 \frac{1}{p^2 - M^2 +i \epsilon} (\bar n \cdot p) \delta (n\cdot p - x M_N) \cr
& = \int \frac{d p^+ dp^- d^2 \vec p_\perp}{2 (2\pi)^4}
  \frac{1}{(p+k)^2 - M^2 +i \epsilon}
 \frac{1}{p^2 - M^2 +i \epsilon} (\bar n \cdot p) \delta (n\cdot p - x M_N). \nonumber
\end{align}

Taking the limit $v \to 1$ before the integral
and then performing the $p^+$ integration with the $\delta-$function,
  we find the following poles on the complex $p^-$ plane.
\begin{align}\label{eq:pintegral_poles_v=1}
p^-_1 &= \frac{k^3(k^3+xM_N) + M_{pk}^2}{k^3 + x M_N}
- i \epsilon \;\mathrm{sign} (k^3 + x M_N), \\
p^-_2 &= \frac{M_p^2}{x M_N} - i \epsilon \;\mathrm{sign} (x).
\end{align}
Here the notation $M_{p\ldots k}^2 = M^2 + (\vec p +\ldots +\vec k)^2 $
is introduced for convenience.
We obtain an important condition for the integral not to vanish,
\begin{align}\label{eq:pintegral_condition_v=1}
\mathrm{sign}(x)\mathrm{sign}(k^3+xM_N)<0.
\end{align}

When we try to integrate before taking $v \to 1$, we find the poles at
\begin{align}\label{eq:pintegral_poles1_v.ne.1}
p^-_{11} &=
\frac{1}{v-1} \left((k_3 + x M_N)v + \sqrt{(k_3+x v M_N)^2+(1-v^2)M^2_{pk}}\right)
+i\epsilon \frac{1+v}{2\sqrt{(k_3+x v M_N)^2+(1-v^2)M^2_{pk}}}\\
\label{eq:pintegral_poles2_v.ne.1}
p^-_{12} &=
\frac{1}{1-v} \left((k_3 + x M_N)v - \sqrt{(k_3+x v M_N)^2+(1-v^2)M^2_{pk}}\right)
-i\epsilon \frac{1+v}{2\sqrt{(k_3+x v M_N)^2+(1-v^2)M^2_{pk}}}\\
\label{eq:pintegral_poles3_v.ne.1}
p^-_{21} &=
\frac{1}{1-v} \left(x M_N v + \sqrt{(x v M_N)^2+(1-v^2)M^2_{p}}\right)
+i\epsilon \frac{1+v}{2\sqrt{(x v M_N)^2+(1-v^2)M^2_{p}}} \\
\label{eq:pintegral_poles4_v.ne.1}
p^-_{22} &=
\frac{1}{1-v} \left(x M_N v - \sqrt{(x v M_N)^2+(1-v^2)M^2_{p}}\right)
-i\epsilon \frac{1+v}{2\sqrt{(x v M_N)^2+(1-v^2)M^2_{p}}}
\end{align}

One important requirement for the integral is that we must recover the
result of the corresponding integral for $v=1$.
This proper limit to the PDF is achieved by closing the contour using
 the upper or half semi-circle, equivalently.
In any case, one of the two poles included in the contour
 approaches to the pole for $v=1$ while the other approaches to infinity and does not
 contribute to the integral in the limit.
Note that two zeros from the same denominator $p_{11}^-$ and $p_{12}^-$(
$p_{21}^-$ and $p_{22}^-$) are always separated by the real axis.
Moreover, the poles do not cross the real axis and thus the requirement
\eqref{eq:pintegral_condition_v=1} is satisfied.

\end{document}